\documentclass[fleqn,twoside]{article}

\usepackage{gc,amssymb,cite}

\eqsection

\newcommand{\be}{\beq}
\newcommand{\ee}{\eeq}

\newcommand{\R}{ \mathbb{R} }

\bls{0.95}
\begin{document}
\twocolumn[

\vspace{-5mm}

\Title
  {Multidimensional cosmology with anisotropic fluid: \yy
  acceleration and variation of $G$ }

\Aunames{J.-M. Alimi\auth{1,a}, V.D. Ivashchuk\auth{2,b,c},
   S.A. Kononogov\auth{3,b} and V.N. Melnikov\auth{4,b,c}}

\Addresses{
 \addr a {Laboratoire de l'Univers et de ses Th\'eories
  CNRS UMR8102,  Observatoire de Paris
  92195, Meudon Cedex, France}
 \addr b {Centre for Gravitation and Fundamental Metrology,
     VNIIMS, 46 Ozyornaya St., Moscow 119361, Russia}
 \addr c {Institute of Gravitation and Cosmology,
     Peoples' Friendship University of Russia,
    6 Miklukho-Maklaya St., Moscow 117198, Russia}
    }


\Abstract
{A multidimensional cosmological model describing the dynamics of $n+1$
 Ricci-flat factor-spaces $M_i$ in the presence of a one-component
 anisotropic fluid is considered. The pressures in all spaces are
 proportional to the density: $p_{i} = w_i \rho$, $i = 0,...,n$. Solutions
 with accelerated expansion of our 3-space $M_0$ and small enough variation
 of the gravitational constant $G$ are found. These solutions exist for two
 branches of the parameter $w_0$. The first branch describes superstiff
 matter with $w_0 > 1$, the second one may contain phantom matter with
 $w_0 < - 1$, e.g., when $G$ grows with time. }


] 
\email 1 {Jean-Michel.Alimi@obspm.fr}
\email 2 {rusgs@phys.msu.ru}
\email 3 {kononogov@vniims.ru}
\email 4 {melnikov@phys.msu.ru}

\section{Introduction}

There are many hot spots in the gravitational interaction \cite{Solv}. Among
them one may point out such problems as acceleration of the Universe
expansion, description and detection of strong field objects (black holes,
wormholes etc.) and gravitational waves, near-zone experiments, such as
equivalence principle tests, second-order tests, rotation and torsion
effects of general relativity etc. Within the last block, we want to stress
a special role of experiments to measure the value of the gravitational
(or Einstein) constant and its possible variations. These experiments
already belong to new generation ones since they are testing not only the
gravitational interaction but also some predictions of unified models and
theories. A special role in these activities is played by space experiments,
and this role will increase in the future. Modern cosmology and its
observational part already became an arena for testing predictions of high
energy physics. All this leads to a fundamental role of gravity in the
present investigations, it is still a missing link in unified theories.
Fundamental physical constants, relations between them and their possible
variations are a reflection of the situation with unification
\cite{Solv,KM,1,6}.

We will be mainly interested in the gravitational constant and its possible
variations. There are three problems connected with the Newtonian
gravitational constant $G$ \cite{Solv}:

 1. Absolute value of $G$.

 2. Possible time variations of $G$.

 3. Possible range variations of $G$, or new interactions (forces).

The oldest one is the problem of possible temporal variation of $G$, which
arose due to papers of Milne (1935) and Dirac (1937). In Russia, these ideas
were developed in the 60s and 70s by K.P.  Staniukovich \cite{S,1}, who
was the first to consider simultaneous variations of several fundamental
constants.

Our first calculations based on general relativity with a perfect fluid and
a conformal scalar field \cite{ZM} gave $\dot{G}/G$ at the level of
$10^{-11} - 10^{-13}$ per year. Our calculations in string-like \cite{IM1}
and multidimensional models with perfect fluid \cite{BIM} gave the level
$10^{-12}$, those based on a general class of scalar-tensor theories
\cite{BMN} and simple multidimensional model with p-branes \cite{IMW,M-G}
gave for the present values of cosmological parameters $10^{-13}-10^{-14}$
and $10^{-13}$ per year, respectively. Similar estimations were made by
Miyazaki within Machian theories \cite{MZ} giving for $\dot{G}/G$ the
estimate $10^{-13}$ per year and by Fujii --- on the level $10^{-14}-
10^{-15}$ per year \cite{FUJ}. Analysis of one more multidimensional model
with two curvatures in different factor spaces gave an estimate on the
level $10^{-12}$ \cite{DIKM}. Here we continue our studies on variation of
$G$ in another multidimensional cosmological model.

\section{The model}

We consider a cosmological model describing the dynamics of $n$ Ricci-flat
spaces in the presence of a 1-component ``perfect-fluid'' matter \cite{IM5}.
The metric of the model
\beq \label{2.1}
      g= - \exp[2{\gamma}(t)]dt \otimes dt +
            \sum_{i=0}^{n} \exp[2{x^{i}}(t)] g^{i}
 \eeq
is defined on the manifold
\beq \label{2.2}
      M = \R \times M_{0} \times \ldots \times M_{n},
\eeq
where $M_{i}$ with the metric $g^{i}$ is a Ricci-flat space
of dimension $d_i$, $i = 0, \ldots ,n $; $n \geq 2$.
The multidimensional Hilbert-Einstein equations have the form
\be \label{2.3}
       R^{M}_{N}-\frac{1}{2}\delta^{M}_{N}R = \kappa^{2}T^{M}_{N},
\ee
where $\kappa^{2}$ is the gravitational constant, and the
energy-momentum tensor is adopted as
 \beq                                           \label{2.6}
    (T^{M}_{N}) = \diag (- \rho,\  p_{0} \delta^{m_{0}}_{k_{0}},
    \ldots ,\ p_{n} \delta^{m_{n}}_{k_{n}}),
 \eeq
describing, in general, an anisotropic fluid.

We put pressures of this ``perfect'' fluid in all spaces to be proportional
to the density,
\beq \label{2.8}
      p_i(t) = (1- u_i/d_i) \rho(t),
\eeq
where $u_i = \const$, $i = 0, \ldots ,n$. We also put $\rho >0$.

We impose also the following restriction on the vector $u = (u_i) \in
 \R^{n+1}$:
\be \label{2.9}
        \aver{u,u}_{*} < 0.
 \ee
Here, the bilinear form $\aver {.,.}_{*}: \R^{n+1} \times \R^{n+1}
 \to \R$ is defined by the relation
  \be      \label{2.10}
        \aver {u,v}_{*} = G^{ij} u_i v_j,
  \ee
 $u,v \in \R^{n+1}$, where
 \beq         \label{2.11}
     G^{ij} = \frac{\delta^{ij}}{d_i} + \frac{1}{2-D}
 \eeq
 are components of the matrix inverse to the matrix of
 the minisuperspace metric \cite{IM2,IMZ}
\beq \label{2.12}
      G_{ij} = d_{i} \delta_{ij} - d_{i} d_{j}.
\eeq
 In (\ref{2.11}), $D = 1 + \sum_{i=0}^{n} d_i$ is the total
 dimension of the manifold $M$ (\ref{2.2}).

  The restriction (\ref{2.9}) reads
\beq  \label{2.13}
   \aver{u,u}_{*}  =   \sum_{i= 0}^{n} \frac{(u_{i})^{2}}{d_{i}}
   + \frac{1}{2-D}\biggl(\sum_{i= 0}^{n} u_{i}\biggr)^{2} < 0.
\eeq

 \section{Solutions with power-law  scale factors}

Here, we consider a special family of ''power-law'' solutions from
\cite{IM5,IM10} with the metric written in the synchronous time
parametrization
\be \label{3.1}
     g=- dt_s \otimes dt_s + \sum_{i= 0}^{n} a_i^2(t_s) g^{i}.
\ee
 Solutions with a power-law behaviour of the scale factors take place for
\be   \label{3.2}
     \aver {u^{(\Lambda)} - u, u}_{*} \neq 0.
\ee
 Here and below the vector
\be  \label{3.3}
         u^{(\Lambda)}_i = 2d_i
\ee
 corresponds to the $\Lambda$-term fluid with $p_i = - \rho$ (vacuumlike
 matter).

 In this case, the solutions are determined by the metric (\ref{3.1})
 with the scale factors
\be   \label{3.4}
       a_i = {a_i}(t_s)  = A_i t_s^{\nu^i},
\ee
and the density
\be   \label{3.4a}
  \kappa^2 \rho = \frac{- 2
    \aver{u, u}_{*}}{\aver{u^{(\Lambda)} - u , u}_{*}^2 t_s^2}.
\ee
Here
\be \label{3.5}
     \nu^i  = 2 u^i/ \aver {u^{(\Lambda)} - u , u}_{*}
\ee
where $u^i =   G^{ij} u_i$ and $A_i$ are positive constants, $i = 0,
\dots, n$.

 The model under consideration was integrated in \cite{IM5} for
 $\aver{u,u}_{*} < 0$. The solutions from \cite{IM5} were generalized in
 \cite{IM10} to the case when a massless minimally coupled scalar field was
 added. Families of exceptional solutions with power-law and exponential
 behaviours of the scale factors in terms of synchronous time were singled
 out in \cite{IM10} and correspond to a constant value of the scalar field:
 $\varphi = \const$.  When the scalar field is omitted, we are led to
 solutions presented above (in \cite{IM5} these solutions were originally
 written in the harmonic time parametrization). It may be verified
 that the exceptional solutions with power-law dependence of scale
 factors are also valid when the restriction (\ref{2.9}) is omitted.
 Moreover, it may be shown that for $\aver{u,u}_{*} = 0$ the power-law
 solutions coincide with the vacuum Kasner-like solution from \cite{I}. In
 this case, the matter source vanishes since $\rho = 0$ in (\ref{3.4a}).

 \section{Acceleration and variation of G}

 In this section, the metric $g^0$ is assumed to be flat, and $d_0 =3$. The
 subspace $(M_0,g^0)$ describes ``our'' 3-dimensional space and
 $(M_i,g^i)$ internal factor-spaces.

 We are interested in solutions with accelerated expansion of our space and
 small enough variations of the gravitational constant obeying the present
 experimental constraints, see \cite{IMW}:
\be \label{4.1v}
        |\dot{G}/(GH)|(t_{s0}) < 0.1,
\ee
 where
\be \label{4.1h}
                H = \frac{\dot{a_0}}{a_0}
\ee
is the Hubble parameter. We suppose that the internal spaces are
compact.  Hence  our 4-dimensional constant is (see \cite{BIM})
\be \label{4.1g}
                G = \const \cdot \prod\nolimits_{i=1}^{n}( a_{i}^{-d_i}).
\ee

  We will use the following explicit formulae for the contravariant
  components:
\be \label{4.2}
   u^i = G^{ij} u_j = \frac{u_i}{d_i} + \frac{1}{2-D} \sum_{j =0}^{n} u_j,
\ee
  and the scalar product reads
\bearr \label{4.3}
   \aver{ u^{(\Lambda)} - u, u }_{*}
\nnn \nq
   =  - \sum_{i= 0}^{n} \frac{(u_{i})^{2}}{d_{i}}
   - \frac{2}{D-2}\sum_{i = 0}^{n} u_{i}
   + \frac{1}{D-2}(\sum_{i= 0}^{n} u_{i})^{2}.
\ear

\subsection{Power-law expansion with acceleration}

 For solutions with power-law expansion, an accelerated
 expansion of our space takes place for
\be \label{4.4}
     \nu^0 > 1.
\ee
 For  $D=4$, when internal spaces are absent, we get
\bearr                                       \label{4.4a}
    \nu^0 = 2/(6 - u_0),
\\ \lal    \label{4.2p}
   \aver {u^{(\Lambda)} -u, u }_{*} = \frac {1}{6}(u_0 -6) u_0 \neq 0,
\ear
 which implies $u_0 \neq 0$ and $u_0 \neq  6$ (here $\aver {u,u}_{*} = -
 \frac{1}{6} u_0^2 < 0$).  The condition  $\nu^0 >1$ is equivalent to $4 <
  u_0 < 6$, or, equivalently,
\beq
      - \rho < p < - \rho/3,
\eeq
  which agrees with the well-known result for $D =4$.
  (We note that recently
  special 5-dimensional power-law solutions (e.g., with acceleration) were
  considered in \cite{PZC}).

  For power-law solutions we get
\be \label{4.5a}
     \frac{\dot{G}}{G} =  - \frac{\sum_{j =1}^{n} \nu^i d_i}{t_s}, \cm
     H = \frac{\dot{a_0}}{a_0} = \frac{\nu^0}{t_s},
\ee
  and hence
\be \label{4.5}
   \dot{G}/(GH) =
    - \frac{1}{\nu^0} \sum_{j =1}^{n} \nu^i d_i \equiv \delta.
\ee
  The constant parameter  $\delta$ describes variation of the gravitational
  constant and, according to  (\ref{4.1v}),
\be \label{4.6}
            |\delta| < 0.1.
\ee

  It follows from the definition of $\nu^i$  in (\ref{3.5}) that
\be \label{4.7}
        \delta = - \frac{1}{u^0} \sum_{i =1}^{n} u^i d_i,
\ee
   or, in terms of covariant components (see (\ref{4.2}))
\be \label{4.8}
    \delta = - \frac{(D-4) u_0 -2 \sum_{i =1}^{n} u_i}{\frac{1}{3} (5 - D)
                u_0 + \sum_{i =1}^{n} u_i}.
\ee

   Thus the relations  (\ref{4.3}), (\ref{4.4}), (\ref{4.6}), (\ref{4.8})
   and the constraint (\ref{3.2}) determine a set of parameters $u_i$
   compatible with the acceleration and tests on G-dot.

   In what follows we will show that these relations do really determine a
   non-empty set of parameters $u_i$ describing the equations of state.

\subsubsection{The case of constant $G$}

   Consider the most important case $\delta =0$, i.e., when the variation
   of $G$ is absent: $\dot{G} = 0$.

   Indeed, there is a tendency of lowering the upper bound on $\dot G$.
   Moreover, according to arguments of \cite{BZhuk}, $\delta < 10^{-4}$.
   This severe constraint just follows from the identity
\be \label{4.Al}
       \dot{G}/G = \dot{\alpha}/\alpha
\ee
   that takes place in multidimensional models. Here $\alpha$ is the fine
   structure constant.

\medskip\noi
 {\bf Isotropic case.}  First we consider the isotropic case when the
 pressures coincide in all internal spaces.  This takes place when
\be  \label{4.9}
        u_i = v d_i, \cm  i = 1, \ldots, n.
\ee
 For pressures in internal spaces we get from  (\ref{2.8})
\be \label{4.10}
      p_i = (1- v) \rho,  \cm i = 1, \ldots, n .
\ee
   Then we get from (\ref{2.13}) and (\ref{4.3})
\bearr   \label{4.11a} \nhq                                  
    \aver{u, u}_{*} = \frac{1}{2-D}
        [- \fract{1}{3} (d-1) u_0 + 2d u_0 v- 2d v^2],
\\ \lal                         \label{4.11b}
    \aver{ u^{(\Lambda)} - u, u }_{*} =
                    \frac{1}{2-D}\bigl[ 2 u_0 + 2dv
\nnn  \cm  \cm
    + \fract{1}{3}(d-1) u_0^2- 2d u_0 v+ 2d v^2 \bigr].
\ear
  Here and in what follows we denote $d = D-4$.

  For $\delta = 0$, we get in the isotropic case
\be  \label{4.12}
        v = u_0/2,
\ee
   or, in terms of pressures,
\be  \label{4.12p}
          p_i= ( 3 p_0 - \rho)/2, \cm  i = 1, \ldots, n.
\ee

   Substituting  (\ref{4.12}) into (\ref{4.11a}) and  (\ref{4.11b})
   we get
\bearr   \label{4.13a}
     \aver{u, u}_{*} = - u_0^2/6,
\\ \lal                     \label{4.13b}
    \aver{u^{(\Lambda)} - u, u }_{*} = u_0 (u_0 -6)/6.
\ear

   Remarkably, we obtain the same relations as in $D = 4$ case (see
   the remark above). For our solution, we should put $u_0 \neq 0$ and $u_0
   \neq 6$.

   Using (\ref{4.9}) and (\ref{4.12}) we get  $u^0 = - u_0/6$ and $u^i = 0$
   for $i > 0$,  hence  $ \nu_i =0$  for $i = 1, \ldots, n$, i.e., all
   internal spaces are static.

   The metric  (\ref{3.1}) reads in our case
\be \label{4.14}
        g=- dt_s \otimes dt_s + A_0^2 t_s^{2 \nu^0} g^{0} +
                \sum_{i= 1}^{n} A_i^2 g^{i},
\ee
   where $A_i$ are positive constants,  and
\be                                                \label{4.15}
   \nu^0 = {2}/(6 - u_0).                          
\ee
   We see that the power $\nu^0$ is the same as in case $D =4$.
   For the density we get from (\ref{3.4a})
\be   \label{4.16}
        \kappa^2 \rho =  \frac{12}{(u_0 -6)^2 t_s^2}.
\ee

  Thus the equations of state (\ref{2.8}), with relations (\ref{4.9})
  and (\ref{4.12}) imposed, lead to the solution (\ref{4.14})--(\ref{4.16})
  with a Ricci-flat (e.g., flat) 3-metric and $n$ static internal Ricci-flat
  spaces. For
\be     \label{4.17}
    4 < u_0 < 6
\ee
  or, equivalently,   $- \rho < p_0 < - \rho/3$, we get an accelerated
  expansion of ``our'' 3-dimensional Ricci-flat space.

\medskip\noi
 {\bf Aisotropic case.} Consider the anisotropic (w.r.t. internal spaces)
 case with $\delta = 0$, or, equivalently (see (\ref{4.8})) ,
\be  \label{4.18a}
        (D-4) u_0 = 2 \sum_{i =1}^{n} u_i.
\ee
   This implies
\bearr  \label{4.18}
     \aver{ u^{(\Lambda)} - u, u }_{*} = \fract{1}{6}u_0 (u_0 -6) - \Delta,
\\ \lal   \label{4.18b}
     \aver{ u, u }_{*} = - \fract{1}{6}  u_0^2 + \Delta,
\ear
where
\be \label{4.19}
  \Delta= \sum_{i=1}^n \frac{u_i^2}{d_i}
   - \frac{1}{d} \biggl(\sum_{i=1}^n  u_i\biggr)^2 \geq 0,  \quad
        d = D - 4.
\ee

 The inequality in (\ref{4.19}) can be readily proved
 using  the well-known  Cauchy-Schwarz inequality:
\be \label{4.19CS}
   \biggl(\sum_{i=1}^n b_i^2\biggr) \biggl(\sum_{i=1}^n c_i^2\biggr)
   \geq \biggl(\sum_{i=1}^n  b_i c_i \biggr)^2.
\ee
  Indeed, substituting $b_i = \sqrt{d_i}$ and  $c_i = u_i/ \sqrt{d_i}$
  into  (\ref{4.19CS}), we get  (\ref{4.19}). The equality in (\ref{4.19CS})
  takes place only when the vectors $(b_i)$ and $(c_i)$ are linearly
  dependent, that for our choice reads:  $u_i/ \sqrt{d_i} = v \sqrt{d_i}$
  where $v$ is constant. Thus $\Delta = 0$ only in the isotropic case
  (\ref{4.9}). In the anisotropic case we get $\Delta > 0$.

  In what follows we will use the relation
\be \label{4.19a}
      \aver { u^{(\Lambda)} - u, u }_{*} =
         \fract{1}{6} (u_0 - u_0^{+})(u_0 - u_0^{-}),
\ee
   where
\be \label{4.19b}
       u_0^{\pm} = 3 \pm \sqrt{9 + 6 \Delta}
\ee
  are roots of the quadratic trinomial (\ref{4.18}) obeying
\be \label{4.19c}
        u_0^{-} < 0, \qquad   u_0^{+} > 6 \qquad
            {\rm for}\quad \Delta > 0.
\ee
    It follows from  (\ref{4.18a}) that $u^0 = - u_0/6$ and hence
\be  \label{4.20}
   \nu^0 = - \frac{2u_0}{u_0 (u_0 -6) - 6\Delta}
\ee
  (here  $u_0 \neq u_0^{\pm}$).

  The function $\nu^0 (u_0)$ is monotonically increasing: (i) from 0 to
  $+\infty$ in the range $(- \infty, u_0^{-})$; (ii) from $-\infty$ to
  $+\infty$ in the range $(u_0^{-}, u_0^{+})$; (iii) from $-\infty$ to  $0$
  in the range $(u_0^{+}, + \infty)$. This behaviour of $\nu^0 (u_0)$
  in each of the three ranges simply follows from the relation
\be  \label{4.21}
   \frac{d \nu^0}{du_0} =  \frac{2 (u_0^2  + 6 \Delta)}
   {(u_0 - u_0^{+})^2(u_0 - u_0^{-})^2}  > 0.
\ee

   The accelerated expansion of our space takes place
   when $\nu^0 > 1$, or, equivalently, when either
\bearr  \label{4.22a}
    {\bf (A)} \quad u_0 \in (u_{0*}^{-}, u_0^{-}), \quad {\rm or}
\\ \lal       \label{4.22b}
    {\bf (B)} \quad u_0 \in (u_{0*}^{+}, u_0^{+}),
\ear
    where
\be  \label{4.23}
           u_{0*}^{\pm} = 2 \pm \sqrt{4 + 6 \Delta}.
\ee
   In terms of the parameter $w_0$,
\be  \label{4.24}
       p_0 = w_0 \rho, \qquad   w_0 = 1 - u_0/3,
\ee
    these two branches read:
\bearr \nq
    {\bf (A)}
\nnn \nq \label{4.25a}
    w_0^{-} = \sqrt{1 + \fract{2}{3} \Delta} < w_0
    < \fract{1}{3} + \fract{2}{3} \sqrt{1 + \fract{3}{2} \Delta}
    = w_{0*}^{-},
\yyy \nq
    {\bf (B)}
\nnn \nq \label{4.25b}
    w_0^{+} = - \sqrt{1 + \fract{2}{3} \Delta} < w_0
    < \fract{1}{3} - \fract{2}{3} \sqrt{1 + \fract{3}{2} \Delta} =
    w_{0*}^{+}.
\ear
     The first branch (A) describes superstiff matter ($w_0 > 1$) with
     negative density. Indeed, $\rho < 0$ follows from (\ref{3.4a}) and
     $\aver{ u,u }_{*} > 0$, see (4.30).

     The second  branch (B) corresponds to matter with a broken weak
     energy condition (since $w_0 < - \frac{1}{3}$) and positive density
     (since $\aver{u, u }_{*} < 0$). This matter is phantom (i.e., $w_0
     < - 1$) when $\Delta \geq 2$.  For $\Delta < 2$ the interval $(w_0^{+},
     w_{0*}^{+})$ contains both ``phantom'' ($w_0 < - 1$) and
     ``non-phantom'' points ($w_0 > - 1$).

\subsubsection{The case of varying $G$}

   Now  we consider another important case $\delta \neq 0$, i.e., when
   $\dot{G} \neq 0$. In what follows, we use the observational bound
   (\ref{4.6}):  $|\delta| < 0.1$,  stating the smallness of  $\delta$.

   Using (\ref{4.8}), we get
\be  \label{5.18a}
            \sum_{i =1}^{n} u_i = \frac{1}{2} d b u_0 ,
\ee
   where $d = D-4$ and
\be  \label{5.18bd}
         b  = b(\delta) = \frac{1+ \delta (1-d)/(3d)}{1 -  \delta/2}.
\ee
   For the scalar product we get from (\ref{5.18a})
\bearr  \label{5.18}
     \aver{ u^{(\Lambda)} - u, u }_{*} =
     \fract{1}{6} A u_0^2  - B u_0  - \Delta,
\\ \lal   \label{5.18b}
     \aver { u, u }_{*} = - \fract{1}{6} A u_0^2 + \Delta,
\ear
  where $\Delta$ was defined in (\ref{4.19}) (see (\ref{2.13}) and
   (\ref{4.3})),
\bear  \label{5.18c}
     \frac{A}{6} \eql \frac{1}{d+2}\biggl(1 + \frac{d}{2}b\biggr)^2 -
                \frac{d}{4}b^2 - \frac{1}{3},
\\  \label{5.18d}                                            
     B \eql \frac{1}{d+2}  (2 + db).
\ear
  Using (\ref{5.18bd}), we obtain the explicit formulae
\bear  \label{5.18cc}
     A \eql A(\delta) = 1 - \frac{(d+2) \delta^2}{12d (1 - \delta/2)^2},
\\  \label{5.18dd}
     B \eql B(\delta) = \frac{1 - \delta/3}{1 - \delta/2}.
\ear
  It should be noted that, due to $|\delta| < 0.1$,  $A$
  is positive,  $A > 0$, and close to unity: $|A -1| < \frac{1}{3} 10^{-2}$.

  For the contravariant component $u^0$ we get from (\ref{4.2}) and
  (\ref{5.18a}):
\be  \label{5.18f}
     u^0 = - C u_0/6,
\ee
   where
\be  \label{5.18ff}
     C = C(\delta) = 3B  - 2  = 1/(1 - \delta/2).
\ee
   It follows from (\ref{5.18b}) and (\ref{5.18f}) that (see (\ref{3.5}))
\be  \label{5.20}
    \nu^0 = - \frac{2 C u_0}{A u_0^2/6  - B u_0  - \Delta}.
\ee
   Here  $u_0 \neq u_0^{\pm}$ where
\be \label{5.19b}
     u_0^{\pm} = u_0^{\pm}(\delta) = \frac{1}{A}
                   (3 B \pm \sqrt{9 B^2  + 6 A \Delta} )
\ee
  are roots of quadratic trinomial (\ref{5.18}).

  In what follows we will use the identity
\be  \label{5.20c}
    \nu^0 -1 = - \frac{A u_0^2  - 4 u_0  - 6 \Delta}
    {A u_0^2  - 6 B u_0  - 6 \Delta}.
\ee

\noi
 {\bf Isotropic case.} Let us consider the isotropic case (\ref{4.9}).
 Then we obtain from  (\ref{5.18a})
\be  \label{5.12}
         v  = d b u_0 /2.
\ee
   or, in terms of pressures
\be  \label{5.12p}
     p_i= \frac{1}{2} [ 3b p_0 + (2 -3b) \rho],\qquad  i = 1, \ldots, n.
\ee
     For scalar products we get
\bearr   \label{5.13a}
     \aver{ u, u}_{*} = - A u_0^2/6,
\\ \lal              \label{5.13b}
    \aver { u^{(\Lambda)} - u, u }_{*} = u_0 (A u_0 - 6 B)/6.
\ear

    For our solution, we should put $u_0 \neq 0$ and $u_0 \neq 6B/A$.
    The metric  (\ref{3.1}) reads in our case
\be \label{5.14}
        g= - dt_s \otimes dt_s +  A_0^2 t_s^{2 \nu^0} g^{0} +
         t_s^{2 \nu} \sum_{i= 1}^{n} A_i^2 g^{i},
\ee
   where $A_i$ are positive constants,
\bear                                                   \label{5.15}
     \nu^0 \eql - \frac{2C}{A u_0 - 6B}, \quad {\rm and}
\\            \label{5.15a}
   \nu \eql  \nu^i = \frac{2 \delta}{d(1 - \delta/2) (A u_0 - 6B)},
\ear
   $i = 1, \ldots, n$.  The last formula follows from (\ref{3.5}) and
\be                                               \label{5.15b}
   u^i =  \frac{u_0 \delta}{6d(1 - \delta/2)}.
\ee
  We see that the power $\nu^0$ does not coincide, for $\delta \neq 0$, with
  that in case $D =4$.

  For the density, since $A >0$, we get from (\ref{3.4a})
\be   \label{5.16}
        \kappa^2 \rho =  \frac{12A}{(Au_0 -6B)^2 t_s^2} > 0.
\ee

  The accelerated expansion condition for our 3D space, $\nu^0 > 1$, reads
  in this case
\be \label{5.17}
    \frac{4}{A(\delta)} < u_0 < \frac{6B(\delta)}{A(\delta)}
\ee
  or, equivalently, in terms of $w_0$  (\ref{4.24}) ($p_0 = w_0 \rho$)
\be \label{5.17a}   \nq\,
     w_{0}^{+}(\delta) =  1 - \frac{2B(\delta)}{A(\delta)}  < w_0 < 1 -
     \frac{4}{3A(\delta)} = w_{0*}^{+}(\delta).
\ee

  For $\delta > 0$, we get an isotropic contraction of the whole internal
  space $M_{1} \times \ldots \times M_{n}$. In this case,
  $w_{0}^{+}(\delta)  < - 1$, and hence phantom matter may occur with the
  equation of state close to the vacuum one since
\be \label{5.17b}
     w_{0}^{+}(\delta) +  1 =  - \frac{\delta (1 + \delta/d)}
     {3[1 - \delta + (d -1) \delta^2/(6d)]}.
\ee
  For small $\delta$ we have $w_{0}^{+}(\delta) =-1 - \delta/3 + O(\delta^2)$.

  For $\delta < 0$ we get an isotropic expansion of the whole internal
  space. Then $w_{0}^{+}(\delta)  > - 1$, and phantom matter does not
  occur. In both cases  $w_{0*}^{+}(\delta) < - 1/3$
  and  $w_{0*}^{+}(\delta) + 1/3 = O(\delta^2)$.

\medskip\noi
  {\bf Anisotropic case.} Now we consider the anisotropic case
  $\Delta > 0$ when $\delta \neq 0$. Using (\ref{5.20c}), we obtain
\be  \label{5.20a}
    \nu^0 -1 = - \frac{(u_0 - u_{0*}^{+})(u_0 - u_{0*}^{-})}
        {(u_0 - u_0^{+})(u_0 - u_0^{-})}.              
\ee
  where $u_0^{\pm} = u_0^{\pm}(\delta)$ were defined in (4.55) and
\be                                               \label{5.19c}
    u_{0*}^{\pm} = u_{0*}^{\pm}(\delta) =
                2 \pm \sqrt{4 + 6 A(\delta) \Delta}.
\ee

Accelerated expansion of our 3-dimensional space takes place
when $\nu^0 > 1$, or, equivalently, when either
\bearr  \label{5.22a}
    {\bf (A)} \quad u_0 \in (u_{0*}^{-}(\delta), u_0^{-}(\delta)),
    \qquad  {\rm or}
\\ \lal    \label{5.22b}
    {\bf (B)} \quad u_0 \in (u_{0*}^{+}(\delta), u_0^{+}(\delta)).
\ear
In terms of the parameter $w_0$ ($p_0 = w_0 \rho$, $w_0 = 1 - \frac{u_0}{3}$)
these two branches read:
\bearr  \label{5.25a}
    {\bf (A)} \quad w_0^{-}(\delta) < w_0  < w_{0*}^{-}(\delta),
\\ \lal         \label{5.25b}
    {\bf (B)} \quad w_0^{+}(\delta)  < w_0  <  w_{0*}^{+}(\delta),
\ear
where
\bear  \label{5.26a}
    w_0^{\pm}(\delta) \eql 1 - {u_0^{\pm}(\delta)}/{3},
\\            \label{5.26b}
       w_{0*}^{\pm}(\delta) \eql 1 - u_{0*}^{\pm}(\delta)/3.
\ear
     For small $\delta$ we have
\bear  \label{5.27a}
  w_0^{\pm}(\delta) \eql w_0^{\pm}(0) -   \frac{\delta}{6}
    \biggl(1 \pm \frac{3}{\sqrt{9 + 6 \Delta}}\biggr) + O(\delta^2),
\\        \label{5.27b}
         w_{0*}^{\pm}(\delta) \eql  w_{0*}^{\pm}(0) + O(\delta^2).
\ear

  Thus for small $\delta$ the lower and upper bounds on $w_0$ have a
  small deviation from those obtained for $\delta = 0$. For small $\delta$,
  the upper bounds shift only by $O(\delta^2)$, while the lower bounds
  shift by $O(\delta)$.

  The first branch (A)  describes superstiff matter $w_0 > 1$ since
  $w_0^{-}(\delta) > 1$ due to  $u_0^{-}(\delta) < 0$.  It may be shown that
  the density is negative in this case since $\aver{u, u }_{*} > 0$.

  For branch (B) we get for the upper bound
\be  \label{5.28a}
            w_{0*}^{+}(\delta) < - 1/3
\ee
  due to $u_{0*}^{+}(\delta) > 4$. For the lower bound we find that
\be  \label{5.28b}
      w_{0}^{+}(\delta) < - 1
\ee
   only if
\be  \label{5.28c}
      \Delta > 6 [A(\delta) -B(\delta)] = - \delta/(1 - \delta/2)^2.
\ee
   This is a condition on the appearance of ``phantom'' matter. For
   $\delta > 0$ this inequality is valid, but for $\delta < 0$ it is
   satisfied only for a big enough value of anisotropy parameter $\Delta$,
   see (\ref{5.28c}).

 \section{Conclusions}

 We have considered multidimensional cosmological models describing the
 dynamics of $n+1$  Ricci-flat factor spaces $M_i$ in the presence of a
 one-component anisotropic fluid with pressures in all spaces proportional
 to the density: $p_{i} = w_i \rho$, $i = 0,...,n$. Solutions with
 accelerated expansion of our 3-dimensional space $M_0$ and small enough
 variation of the gravitational constant $G$ were found. These solutions
 exist for two branches of the parameter $w_0$: (A) from (\ref{5.25a})
 and (B) from (4.74). Branch (A) describes superstiff matter with
 $w_0 > 1$ while branch (B) may contain phantom matter with $w_0 < - 1$. The
 second branch obviously contains phantom matter states when (i)
 $\delta > 0$, i.e., $G$ increases with time, or (ii) $\delta = 0$ and
 $\Delta >0$, i.e., if $G$ is constant and the expansion (or contraction) of
 the factor spaces is anisotropic. If (iii) $\delta < 0$, i.e., $G$
 decreases with time, phantom matter appears if the anisotropy parameter
 $\Delta$ is big enough (see (\ref{5.28c}).)

\Acknow{The  work of V.D.I. and  V.N.M. was supported in part
   by the  DFG grant  Nr. 436 RUS 113/807/0-1 and also
   by the Russian Foundation for   Basic Research, grant Nr.
   05-02-17478.}

 \end{document}